# Groove-shaped defects in as-grown (001)-oriented β-$Ga_2O_3$ epilayers prepared by halide vapor phase epitaxy

Yongzhao Yao,[1,2,a)] Daiki Katsube,[2] Hirotaka Yamaguchi,[2] Yukari Ishikawa[2]

[1]Mie University, 1577 Kurimamachiya-cho, Tsu, Mie 514-8507, Japan
[2]Japan Fine Ceramics Center, 2-4-1 Mutsuno, Atsuta, Nagoya 456-8587, Japan

**Abstract:** Groove-shaped defects (GSDs) degrade the surface flatness of as-grown (001)-oriented β-$Ga_2O_3$ epilayers and necessitate chemical mechanical polishing before device fabrication, increasing processing costs and the risk of damage. We investigated the morphology, subsurface structure, and formation mechanism of GSDs in a homoepitaxial layer grown by halide vapor phase epitaxy using synchrotron X-ray topography and electron microscopy. The GSDs extended several millimeters along [010] and consisted predominantly of ($\bar{1}02$) basal facets bounded by steep (100) sidewalls. Their wafer-scale distribution showed a spatial correspondence with variations in wafer curvature, suggesting that local surface orientation influences GSD formation. Careful alignment of surface images with transmission X-ray topographs revealed no one-to-one correspondence between GSDs and substrate defects, providing no evidence that substrate dislocations serve as their nucleation sites. Instead, transmission electron microscopy revealed planar defects localized near the terminal boundaries where the faceted GSD sectors met the surrounding (001) growth region; no such defects were

a)Author to whom correspondence should be addressed. Electronic mail: yao@icsdf.mie-u.ac.jp.
ORCID:0000-0002-7746-4204

observed in specimens extracted from the middle of GSDs. These defects exhibited α-fringe contrast characteristic of inclined translational planar defects, with the dominant segments assigned to the $(1\bar{2}1)$ plane. The observations suggest that variations in local surface orientation and step supply may promote three-dimensional faceted growth, producing persistent $(\bar{1}02)/(100)$ sectors. The localized planar defects are therefore interpreted as consequences of growth-sector coalescence rather than the origins of GSD nucleation. These findings provide insight into the roles of wafer curvature and surface step supply in GSD formation.

## I. INTRODUCTION

β-$Ga_2O_3$ has emerged as a promising semiconductor for next-generation power electronic devices owing to its ultra-wide bandgap of approximately 4.8–4.9 eV, high critical electric field exceeding 8 MV $cm^{-1}$, and the availability of large-area single-crystal substrates grown from the melt [1, 2, 3]. These advantages enable a Baliga's figure of merit that is significantly higher than those of conventional wide-bandgap semiconductors such as SiC and GaN, making β-$Ga_2O_3$ an attractive material for high-voltage and high-efficiency power devices [4, 5]. For realizing high-performance β-$Ga_2O_3$ power devices, the control of crystal defects is critically important because structural defects can act as carrier trapping centers, leakage current paths, and premature breakdown sites, thereby degrading device performance and reliability [6, 7, 8, 9, 10]. In particular, homoepitaxial layers serve as the active regions where device structures are fabricated. Therefore, defects introduced into the epitaxial layer directly affect device characteristics and long-term operational stability [11, 12, 13].

Various techniques have been employed for β-$Ga_2O_3$ homoepitaxy, including metal-organic vapor phase epitaxy [13, 14, 15, 16], molecular beam epitaxy [17, 18], mist chemical vapor deposition [19], and halide vapor phase epitaxy (HVPE) [11, 20, 21, 22]. Among these methods, HVPE has attracted considerable attention because of its high growth rate, high material utilization efficiency, and suitability for producing thick epitaxial layers required for vertical power device applications. The (001) orientation is one of the most widely used substrate orientations for β-$Ga_2O_3$ epitaxy [10, 18, 23]. A major advantage of this orientation is the availability of large-area substrates fabricated by the edge-defined film-fed growth (EFG) method [24]. Furthermore, because the crystallographic *b*-axis lies within the substrate plane, the propagation of *b*-axis dislocations and pipe-like nanovoids from the substrate into the epilayer is suppressed [6, 25, 26, 27, 28, 29, 30, 31]. In addition, the (001) orientation generally allows relatively high growth rates, facilitating the fabrication of thick epitaxial layers within practical growth times.

Despite these advantages, defect formation in (001)-oriented β-$Ga_2O_3$ epitaxial layers remains a significant challenge. Among the characteristic defects are groove-shaped defects (GSDs), which are frequently observed on the surfaces of as-grown homoepitaxial layers prepared by HVPE [32, 33, 34]. These defects are not merely surface morphological irregularities but are often accompanied by structural defects extending into the epitaxial layer. Their presence degrades surface flatness and prevents the direct use of as-grown epilayers for device fabrication. Consequently, chemical mechanical polishing (CMP) is generally required prior to device processing, increasing manufacturing costs and introducing a risk of additional processing-induced damage [34]. Several studies have investigated the origin of these defects. Goto *et al*. [32] reported streaky grooves and triangular pits on HVPE-grown β-$Ga_2O_3$ layers and showed by transmission electron microscopy (TEM) that the deepest region of the pits is associated with dislocations propagating from the substrate. They proposed that pit formation originates from substrate dislocations intersecting the growth surface. More recently, Sdoeung *et al*. [33] reported that deep pits, which correspond to the GSDs investigated in the present study, spatially correlate with dislocation clusters identified in the substrate by synchrotron X-ray topography (XRT), and suggested that these clusters play an important role in pit formation. Although these studies demonstrated correlations between GSDs and substrate defects, the fundamental mechanism responsible for the formation of GSDs has not been clarified. In contrast to these two reports, Lin *et al*. reported that nonuniform surface regions characterized by undulations and GSDs preferentially formed on substrates with near-zero miscut angles. They attributed the formation of these regions to differences in the epitaxial growth mode, suggesting that near-on-axis substrates promote three-dimensional island growth, whereas the introduction of an appropriate miscut favors step-flow growth and suppresses the development of nonuniform surface morphology. In other words, Lin *et al*. attributed the formation of GSDs to variations in the local substrate miscut angle rather than to crystallographic defects in the substrate [34]. It remains unclear whether the GSDs are directly generated by substrate defects or arise from growth-related processes during epitaxy. A detailed understanding of their geometrical evolution and structural nature is therefore still lacking.

In this work, we systematically investigate the distribution, morphology, and structural characteristics of GSDs in HVPE-grown 2-inch β-$Ga_2O_3$ homoepitaxial layers on (001)-oriented substrates. Laser microscopy, scanning electron microscopy (SEM), X-ray diffraction (XRD), XRT, focused ion beam (FIB) and TEM are employed to reveal their geometrical and crystallographic features. Our results demonstrate that the GSDs originate from the coalescence of growth sectors with different growth orientations, resulting in structural mismatch at their boundaries. In contrast to previous reports [32, 33], we found no evidence that substrate defects directly act as nucleation sites for GSDs.

## II. EXPERIMENTAL DETAILS

A 2-inch-diameter, (001)-oriented substrate was used for HVPE homoepitaxial growth. The substrate was fabricated by EFG method at Novel Crystal Technology, Inc. (NCT). It was 650 μm thick and was Sn-doped to a net donor concentration of $6.3\times10^{18}$ $cm^{-3}$. Before epitaxial growth, the substrate surface was treated by CMP. XRD measurements revealed miscut angles of 0.03° and 0.09° toward the $[\bar{1}00]$ and $[0\bar{1}0]$ directions, respectively. HVPE growth was also performed at NCT under conditions similar to those reported in Refs. 33 and 34. The growth temperature was 1000 °C, and the growth rate was approximately 5 μm $h^{-1}$. The Si dopant concentration was $1.5\times10^{17}$ $cm^{-3}$. The as-grown epilayer was 23 μm thick.

The surface morphology of the as-grown HVPE epilayer was examined using laser microscopy and SEM. XRT was performed in transmission geometry at beamline BL24XU of SPring-8. Transmission XRT utilizes the dynamical X-ray diffraction phenomenon known as the Borrmann effect, or anomalous transmission [35, 36]. Our group has applied this technique to visualize dislocations and other lattice defects in highly X-ray-absorbing crystals, including β-$Ga_2O_3$ [37, 38, 39, 40] and GaN [41, 42]. The wavelength of the monochromatic X-ray beam was set to 0.124 nm. XRT images were acquired using a reciprocal lattice vector of $\boldsymbol{g}$ = 020 at a Bragg angle of 24.1°. The X-rays entered the sample from the substrate side and exited from the epilayer side, allowing lattice defects in both the substrate and epilayer to be imaged. We carefully aligned the transmission XRT with the corresponding surface morphology images. This spatial correlation allowed us to determine whether substrate defects were located directly beneath the GSDs and could therefore serve as their nucleation sites.

Cross-sectional specimens were prepared using FIB milling for SEM and TEM observations. SEM was used to determine the crystallographic orientations of the facets constituting the GSDs. TEM observations were performed at an accelerating voltage of 200 kV to characterize lattice defects beneath the GSDs.

## III. RESULTS AND DISCUSSION

### A. Surface morphology and faceted growth behavior of GSDs

Figure 1 shows the surface morphology of the entire as-grown epilayer. Magnified images of ten representative areas selected from the left (L), right (R), upper (U), and central (C) regions are shown in Figure 2. Based on the image contrast, the surface morphology in Figure 1 can be broadly classified into three groups, labeled (i), (ii), and (iii). The darkest regions, labeled (i), appeared predominantly on the left side of the wafer as lines extending along the [010] direction (e.g., L2). These features correspond to GSDs. Regions exhibiting intermediate contrast, labeled (ii), appeared near the horizontal center of the wafer (e.g., U) and occupied most of the wafer surface. These regions were densely covered with facets inclined at small angles to the (001) surface. Here, "small-angle facets" refers to facets that are too shallow to be classified as GSDs. The GSD density was substantially lower in these regions. The bright regions, labeled (iii), appeared on the right side of the wafer (e.g., R3) and exhibited much lower densities of both GSDs and small-angle facets. Consequently, a larger proportion of the exposed (001) surface resulted in brighter contrast. The boundary between regions (ii) and (iii) formed an arc, as indicated by the green dashed line in Figure 1. As suggested by Lin *et al*. [**34**], this boundary corresponds to locations at which the miscut angle along the [010] direction reaches a particular value. Notably, the termination boundaries of groups of GSDs also exhibited linear or arc-like profiles, as indicated by the yellow dashed lines in Figures 2(b)–2(d). This observation suggests that GSD formation is also associated with the local miscut angle. An unannotated laser microscopy image of the entire wafer is provided in Supplementary Figure S1.

Figure 3(a) shows a detailed SEM image of the surface morphology of GSDs in area L3, and magnified images of representative regions are shown in Figures 3(b)–3(e). The GSDs extended several millimeters along the [010] direction. At their terminal ends, they intersected the surrounding (001) growth surface, forming roughly triangular boundaries, as indicated by the red arrows. Rather than exhibiting well-defined straight edges, these boundaries had irregular, zigzag outlines [Figures 3(c) and 3(e)]. Another notable feature is that neighboring GSDs tended to terminate at similar positions along the [010] direction, collectively forming stepwise or zigzag termination fronts [Figure 3(a)]. Such spatially correlated termination is difficult to explain solely by the independent nucleation of individual GSDs at randomly distributed substrate defects. Instead, it suggests that the propagation and termination of the GSDs were governed by a common local growth condition, such as variations in the substrate miscut angle or surface step density.

Cross-sectional SEM observations further revealed that the interior of each GSD consisted of two crystallographic facets: a relatively flat ($\bar{1}02$) basal facet and a steep (100) sidewall, both of which were smooth and exhibited no discernible secondary faceting. In contrast, the nominal (001) surface surrounding the GSDs was densely covered with small-angle facets. This abrupt morphological transition indicates that the GSDs represent a locally distinct growth mode rather than simple depressions formed on an otherwise uniform growth surface. The persistence of the smooth ($\bar{1}02$) and (100) facets suggests that these planes may have lower normal growth rates than the (001) surface under the present HVPE conditions, whereas the pronounced elongation of the GSDs along [010] suggests strongly anisotropic surface growth kinetics.

Figure 4(a) shows a densely populated group of GSDs in area L2. Two representative GSDs were selected for detailed analysis. P1 and P3 correspond to the terminal regions of the respective GSDs, where the ($\bar{1}02$)-faceted growth regions intersect the surrounding (001) growth surface. These positions therefore correspond to the boundaries described in Figure 3. In contrast, P2 is located within the elongated ($\bar{1}02$)-faceted region, away from the terminal boundary at P1, although both positions lie within the same GSD. Figure 4(b) shows the location at the terminal region P1 from which a cross-sectional specimen was extracted by FIB. To ensure accurate extraction from the center of the terminal boundary, four linear FIB markers were fabricated around P1 in a crosshair-like arrangement. Another cross-sectional specimen was extracted from P2 using the same procedure. Each of the two specimens was examined from two opposite sides, with the viewing directions toward [010] and along [010], as indicated by the blue and red arrows, respectively.

Because P2 was extracted from the middle of a long GSD, the cross-sectional profiles observed toward and along [010] exhibited mirror-related morphologies [Figures 4(c) and 4(d)]. Both views clearly revealed that the GSD consisted predominantly of alternating $(\bar{1}02)$ facets and (100) sidewalls, consistent with the surface observations shown in Figure 3. An additional facet indexed as (701) was locally observed where the GSD intersected the surrounding (001) surface along the $[\bar{1}00]$ direction. Nevertheless, the GSD morphology was dominated by the $(\bar{1}02)$ and (100) facets, indicating that these two planes constitute the fundamental structural units of the GSD, whereas the (701) facet likely serves as a local transition or accommodation facet at the lateral boundary. In contrast to P2, the terminal specimen P1 exhibited a dependence on the viewing direction. The $(\bar{1}02)$ facet and (100) sidewall were indistinct when viewed from the surrounding (001) growth region but became clearly visible when viewed from the GSD side [Figures 4(e) and 4(f), respectively]. This directional asymmetry is characteristic of the three-dimensional termination of the faceted GSD region at its boundary with the surrounding (001) growth region. The termination appears to occur when the surrounding (001) growth front overtakes or closes the faceted region.

P3 also corresponds to the terminal region of a GSD. However, compared with P1, the FIB extraction position at P3 was shifted toward the surrounding (001) growth region to examine its characteristic growth morphology [Figure 4(g)]. The surface exhibited shallow, small-angle facets nearly parallel to the nominal (001) surface [Figure 4(h)]. These facets are therefore expected to have normal growth rates similar to that of the (001) surface. When viewed from the opposite side along [010], a small remnant of the GSD structure became visible at the position indicated by the green arrow [Figure 4(i)].

These observations suggest that the GSD region, dominated by $(\bar{1}02)$-facet and (100)-sidewall growth, impinges on the surrounding (001) growth region, which is dominated by small-angle-facet growth, thereby forming the irregular, zigzag termination boundaries observed in the plan-view SEM images in Figure 3.

Similar $(\bar{1}02)$ and (100) facets were previously observed by Lin *et al*. [34]. They related the stabilization of these facets to anisotropic surface energetics and further suggested that the formation of the undulated regions was associated with a three-dimensional growth mode favored by substrates with near-zero miscut angles. The predominance of the same two facets in the present GSDs suggests that GSD formation is likewise governed by facet-dependent growth kinetics.

Further insight into the role of the $(\bar{1}02)$ facet can be obtained from studies of homoepitaxial growth on $(\bar{1}02)$-oriented β-$Ga_2O_3$ substrates. Oshima and Oshima [43] demonstrated HVPE growth on a native $(\bar{1}02)$ substrate at a rate of 23 μm $h^{-1}$, which was only slightly lower than the 26 μm $h^{-1}$ obtained simultaneously on a (001) substrate. More recently, Sdoeung *et al*. demonstrated HVPE homoepitaxial growth on a 2-inch $(\bar{1}02)$-oriented substrate at 1000 °C with a growth rate of approximately 4.2 μm $h^{-1}$, again comparable to that obtained on the (001) surface under similar growth conditions [44]. In both studies, the resulting epilayers retained the high crystallinity of the substrates, suggesting that the $(\bar{1}02)$ surface can function as a persistent, crystallographically coherent growth front during HVPE rather than merely appearing as a transient facet.

In the present (001)-oriented epilayer, the extended $(\bar{1}02)$ basal facets within the GSDs can therefore be interpreted as locally reoriented growth fronts. Because the normal growth rates of the $(\bar{1}02)$ and (001) surfaces can be comparable under HVPE conditions, the two growth regions may coexist over extended distances without the $(\bar{1}02)$-faceted region being rapidly eliminated by planarization. Such kinetic competition provides a plausible explanation for the millimeter-scale propagation of the GSDs. Moreover, previous growth experiments on ($\bar{1}02$) substrates revealed hillocks and macrosteps elongated along [010] [43], indicating pronounced in-plane morphological anisotropy [45] despite the high normal growth rate of this orientation. This behavior is consistent with the strong [010] elongation of the GSDs observed here.

The consistent association of the ($\bar{1}02$) basal facets with smooth (100) sidewalls provides additional evidence that the GSD morphology is crystallographically selected rather than generated by random surface roughening. The suitability of the ($\bar{1}02$) orientation for forming structures with smooth (100) sidewalls has also been recognized in the context of plasma-free fabrication of vertical fins and trenches [46, 47]. Because the ($\bar{1}02$) and (100) planes form a steep facet junction, their repeated pairing naturally produces the characteristic step-like cross section of the GSDs, consisting of relatively broad ($\bar{1}02$) basal facets separated by steep (100) sidewalls. Once a local ($\bar{1}02$) growth front is established, the low-energy (100) plane may act as a crystallographically preferred sidewall connecting adjacent ($\bar{1}02$) facets and accommodating the transition toward the surrounding (001)-oriented growth region.

Nevertheless, the emergence of the ($\bar{1}02$)/(100) facet pair cannot be explained solely by the preferential exposure of low-energy surfaces [48, 49, 50, 51, 52]. Xie *et al.* calculated surface energies of 0.49, 1.79, and 1.92 J $m^{-2}$ for the (100), (010), and ($\bar{1}02$) planes, respectively [53]. They further reported the following order of surface energies: $\gamma(\bar{1}02) > \gamma(010) > \gamma(001) > \gamma(\bar{2}01) > \gamma(100)$, and suggested that this anisotropy may result in plane-dependent differences in surface bond strengths and diffusion lengths, which in turn affect the $Ga_2O$ desorption process [53]. Thus, whereas the particularly low surface energy of the (100) plane may contribute to the smoothness and persistence of the sidewalls, the $(\bar{1}02)$ plane has a substantially higher surface energy. Oshima further reported that the TMAH etching rate of $(\bar{1}02)$ surface was approximately one order of magnitude higher than those of (100), (010), (001), and $(\bar{2}01)$ surfaces, attributing this high reactivity partly to the high surface energy, charge-neutral atomic configuration, and weak surface rumpling of the $(\bar{1}02)$ plane [46]. Although wet-etching kinetics cannot be directly equated with HVPE growth kinetics, these findings indicate that the $(\bar{1}02)$ surface possesses distinctive reaction kinetics.

### B. XRT analysis of the spatial correlation between GSDs and subsurface defects

Figure 5 compares the surface positions of the GSDs with lattice defects observed by transmission XRT. Figure 5(a) shows a photograph of the strip-shaped specimen extracted from the 2-inch-diameter wafer and a laser microscopy overview image of its surface. The specimen was mounted on a rectangular metal frame for XRT observation. Figure 5(b) shows a transmission XRT image of the area between the yellow dashed lines, obtained with $\boldsymbol{g} = 0\bar{2}0$, whereas Figure 5(c) shows the surface morphology of the same area observed by laser microscopy. The dimensions of the two images were carefully matched to investigate the spatial correlation between the GSDs and lattice defects. The circles indicate corresponding locations in the two images. One representative region, marked by the green circle, was magnified for more detailed comparison in Figures 5(d) and 5(e).

In transmission XRT, dislocations and other extended defects locally suppress anomalous transmission and thereby increase X-ray absorption. Consequently, these defects appear as dark line contrasts in the XRT image [37, 38]. According to dynamical diffraction theory, defects close to the exit surface, corresponding to the epilayer side in the present geometry, produce relatively sharp line contrasts, whereas those close to the entrance surface, corresponding to the substrate side, produce broader and more diffuse contrasts. As shown in Figure 5(b), numerous line-like contrasts originating from defects in both the substrate and the epilayer were visible in the transmission XRT image. Nevertheless, none of these contrasts exhibited a clear one-to-one spatial correspondence with the GSDs identified by laser microscopy. It should be noted that the XRT measurements were performed on the as-grown epilayer without CMP. Deep GSDs can therefore affect the transmitted X-ray intensity and produce faint morphological shadows in the XRT image. These shadows can be distinguished from the diffraction contrast produced by actual lattice defects in the substrate and epilayer based on their contrast characteristics.

In the enlarged region, no distinct defect-related XRT contrast was detected directly beneath the GSDs or at their terminals [Figures 5(d) and 5(e)]. Within the spatial resolution and contrast sensitivity of the present XRT measurements, these observations provide no evidence that pre-existing substrate defects act as nucleation sites for GSD formation or propagate from the substrate into the epilayer beneath the entire GSD. Instead, the XRT results indicate that GSDs are primarily surface morphological features produced during epitaxial growth. It is noted that the absence of distinct defect-related XRT contrast does not necessarily indicate that the GSDs are completely free of internal lattice defects. As demonstrated later by cross-sectional TEM, a localized lattice defect was present within the epilayer at the boundary where the ($\bar{1}02$)-faceted growth region intersected the surrounding (001) growth region. Importantly, this defect was confined to the boundary and did not extend beneath the entire GSD. This result suggests that the localized defect was generated by the impingement of the two growth regions rather than serving as the nucleation origin of the GSD.

### C. Lattice defects at GSD termination boundaries examined by TEM

Figures 6(a)–6(c) show plan-view images acquired during FIB preparation of cross-sectional TEM specimens from the terminal boundaries of three GSDs, designated P4, P5, and P6, respectively. All three GSDs were selected from area L2. The red dashed lines outline the GSDs, while the green arrows indicate the linear FIB markers used to locate the terminal boundaries accurately. The regions enclosed by the yellow dashed lines indicate the positions from which the TEM specimens were extracted. All three specimens were observed along [010], as indicated by the red arrow in Figure 6(a).

Figures 6(d)–6(f) show stitched, low-magnification cross-sectional TEM images of P4, P5, and P6, respectively. Each image was assembled from multiple micrographs to show the entire extent of the defect beneath the corresponding GSD terminal boundary. As described above, the surface profiles at these boundaries consisted primarily of a steep (100) sidewall and a relatively flat ($\bar{1}02$) basal facet. Although these facets are not clearly resolved in Figure 6(d), their original morphology at P4 was confirmed by cross-sectional SEM before final FIB thinning, as shown in Supplementary Figure S2. The white dashed lines in Supplementary Figure S2(b) trace the successive ($\bar{1}02$), (100), and ($\bar{1}02$) facets, and the red arrow indicates the corner selected for TEM observation.

The selected-area electron diffraction patterns shown in the lower-left corners of Figures 6(d)–6(f) exhibit a single set of diffraction spots corresponding to the same crystallographic orientation. No additional diffraction spots attributable to a secondary phase or a strongly misoriented crystalline domain were detected. In all three specimens, a distinct defect originated at the corner where the (100) sidewall met the ($\bar{1}02$) basal facet and extended downward into the epilayer. The defects observed at P4 and P5 exhibited similar elongated, band-like morphologies with several changes in direction [Figures 6(d) and 6(e)]. In contrast, the defect at P6 was considerably shorter and exhibited an approximately parallelogram-shaped morphology [Figure 6(f)]. Figures 6(g)–6(i) schematically illustrate the cross-sectional TEM observations of P4, P5, and P6, respectively. The red dashed lines outline the defects observed beneath the terminal boundaries.

The reproducible occurrence of an internal defect at the terminal boundary of each of the three independent GSDs indicates that these defects are intrinsically associated with GSD termination. In each case, the defect originated at the surface corner between the (100) and $(\bar{1}02)$ facets and extended into the epilayer, with no evidence of propagation from the underlying substrate. This spatial relationship suggests that the defects were generated when the $(\bar{1}02)$-faceted GSD growth region impinged on the surrounding (001) growth region. The coalescence of these crystallographically distinct growth fronts may produce a local displacement or stacking mismatch that is accommodated by the formation of a planar defect extending into the epilayer.

Electron diffraction patterns were also acquired from the surrounding defect-free matrix and from several characteristic positions within the band-like defects, including portions before and after changes in direction, and kinked or corner regions. All the examined regions exhibited a single set of diffraction spots corresponding to the same crystallographic orientation. No second set of diffraction spots or spot splitting corresponding to a mirror-related orientation was observed. Therefore, under the examined diffraction conditions, the defects are unlikely to be large-angle domain boundaries or twin boundaries, although twin formation through double positioning of the adatoms has been reported to occur readily during growth on (100)-oriented surfaces [**14**, **47**, **49**, **52**, **54**, **55**, **56**, **57**, **58**, **59**], which correspond to the GSD sidewalls in the present study. However, these observations do not exclude translational planar defects, such as stacking faults or antiphase boundaries, because such defects can preserve the overall crystallographic orientation of the surrounding matrix. The differences in the apparent lengths and shapes of the defects at P4–P6 may reflect differences in their three-dimensional geometries and in the angles at which they intersect the TEM specimens. These possibilities are examined through the detailed diffraction-contrast analyses presented in Figures 7 and 8, and Supplementary Figure S3.

Figure 7(a) shows a bright-field TEM image of the defect at P4 recorded under a two-beam condition with the 000 and 002 reflections strongly excited. The defect exhibited a dark, band-like contrast extending from the GSD terminal boundary into the epilayer. Figure 7(b) shows the corresponding dark-field image obtained using the same reflection. Magnified images of characteristic regions of the defect are shown in Figures 7(c)–7(e). The narrow line contrasts crossing the bright- and dark-field images are extinction contours caused by slight local bending of the TEM foil and should not be confused with the defect.

The band-like defect consisted of several approximately parallel bright and dark fringes. Such contrast is commonly referred to as $\alpha$-fringe contrast and is characteristic of inclined planar defects such as stacking faults [60]. It is worth noting that the low-magnification TEM images in Figure 6 were acquired with the incident electron beam parallel to the [010] zone axis (see diffraction patterns in Figures 6(d)–6(f)), under which the fringe contrast associated with the planar defects was not clearly resolved. In contrast, the TEM images in Figure 7 were acquired under a two-beam diffraction condition, in which a translational displacement $\boldsymbol{R}$ across a planar defect introduces a phase shift $\alpha = 2\pi\boldsymbol{g}\cdot\boldsymbol{R}$. When the defect plane is inclined with respect to the TEM specimen surfaces, the depth at which the electron beam intersects the defect varies across the image, producing a series of bright and dark fringes. The observed fringe contrast therefore provides strong evidence that the band-like feature is a translational planar defect inclined to the specimen surfaces.

The defect trace changed direction abruptly at several positions [Figures 7(c) and 7(d)], while the fringe packet remained substantially continuous across the bends. This morphology suggests that the defect does not lie on a single crystallographic plane but instead has a folded or faceted three-dimensional geometry consisting of planar segments on different habit planes. These segments may form a single continuous translational boundary joined along fold lines. This interpretation is consistent with the plan-view SEM observations in Figure 3, which showed that the boundary where the $(\bar{1}02)$-faceted GSD growth region impinged on the surrounding (001) growth region had an irregular, zigzag outline rather than forming a single smooth interface. The subsurface defect may therefore inherit the faceted geometry of this growth-front impingement boundary. The local discontinuity and intersection-like contrast in Figure 7(e) may arise from the intersection or overlap of multiple planar-defect segments along the electron-beam direction. Similar contrast has been reported for overlapping planar defects [60, 61]. However, the precise origin of this feature cannot be determined from the present image alone.

Figure 8 presents diffraction-contrast TEM observations of the defect beneath the GSD terminal boundary at P5. As observed at P4, the defect originated from the surface corner where the (100) sidewall met the $(\bar{1}02)$ basal facet and extended downward into the epilayer. Figures 8(a) and 8(b) show overview images obtained under diffraction conditions with $\boldsymbol{g}$ = 002 and $\boldsymbol{g}$ = 400, respectively. The defect exhibited clear contrast under both conditions. Unlike the defect at P4, which appeared predominantly as a single continuous band, the defect at P5 was divided into several mutually offset, approximately parallelogram-shaped fringe packets.

Figures 8(c) and 8(d) show magnified images of the central part of the defect obtained with $\boldsymbol{g}$ = 002 and $\boldsymbol{g}$ = 400, respectively. Each fringe packet consisted of multiple, approximately parallel bright and dark fringes characteristic of an inclined translational planar defect. The visibility of the defect under both diffraction conditions suggests that neither reflection satisfies the ideal invisibility condition, $\boldsymbol{g}\cdot\boldsymbol{R}$ = integer, for the displacement vector $\boldsymbol{R}$. However, determination of R requires systematic observations using additional $\boldsymbol{g}$ vectors.

More fringes appear to be resolved with $\boldsymbol{g}$ = 002 [Figure 8(c)] than with $\boldsymbol{g}$ = 400 [Figure 8(d)]. This difference may partly reflect the dependence of the phase shift on $\boldsymbol{g}$. Nevertheless, the number and spacing of the fringes also depend on the extinction distance, deviation parameter, specimen thickness, and inclination of the defect plane. In addition, different specimen tilts were required to establish the $\boldsymbol{g}$ = 002 and $\boldsymbol{g}$ = 400 diffraction conditions. The resulting change in projection geometry accounts for at least part of the apparent differences in the separations and projected areas of the offset parallelogram-shaped segments. These observations support a three-dimensional configuration consisting of several displaced or faceted planar-defect segments rather than a single planar defect lying on one continuous plane. The offset parallelogram-shaped fringe packets resemble the ribbon-folded defects reported by Ogawa *et al*. in HVPE-grown β-$Ga_2O_3$ [**62**]. They attributed those defects to repeated stacking faults arranged on alternating crystallographic planes, including $(111)/(7\bar{2}7)$ or $(1\bar{1}1)/(727)$. The characteristic dimensions of the segments observed at P5 are approximately one order of magnitude larger than those reported by Ogawa *et al*., indicating that the defects are not necessarily identical. Nevertheless, their morphological similarity suggests that both structures may be produced by the faceting or folding of translational planar defects.

Figure 9 schematically illustrates the three-dimensional geometry of the planar defects inferred from Figures 7 and 8. The longest band-like contrasts in both figures extended along $[\bar{1}01]$, which corresponds to the intersection of a $(1n1)$ plane with the (010) surface of the FIB lamella. As shown in Figure 9(a), the lamella thickness and projected defect width were approximately 170 and 545 nm, respectively. Their geometrical relationship yielded an inclination angle of approximately 72.7° between the defect plane and the (010) surface, consistent with a $(1\bar{2}1)$ habit plane. The bright–dark sequence of the α-fringes indicated that the lowest and uppermost fringes in the TEM image corresponded to the intersections of the defect with the top and bottom surfaces of the lamella, respectively [60]. Figure 9(b) compares the calculated trace of the $(1\bar{2}1)$ plane on the (010) surface, viewed along [010], with the experimentally observed fringe at P4. The close agreement between these directions supports the assignment of the longest planar-defect segments observed in Figures 7 and 8 to the $(1\bar{2}1)$ plane. Supplementary Figure S3 presents detailed TEM observations of P6. Although the defect was shorter than those at P4 and P5, its longest fringe packet extended along $[\bar{1}01]$ and had a comparable projected width. These similarities indicate that the dominant planar-defect segment at P6 also lies on the $(1\bar{2}1)$ plane. Its shorter extent may be attributed to the observation position being slightly farther from the boundary between the $(\bar{1}02)$ and (001) growth regions. Because the results are consistent with those for P4 and P5, further detailed discussion is omitted. Notably, the planar defects were observed only near the GSD termination boundaries. No such defects were detected by TEM in specimens extracted from the middle portions of the GSDs, as demonstrated by the two representative examples in Supplementary Figure S4.

### D. Proposed formation mechanism of GSDs

Based on the above observations, we propose that the local effective miscut angle is a key parameter governing GSD formation. Wafer curvature changes the local normal direction of the (001) lattice plane and therefore produces spatial variations in the effective miscut. Consistent with this interpretation, the XRD curvature map in Supplementary Figure S5, obtained using the previously reported method [63, 64, 65], shows that the red region near the secondary orientation flat spatially coincides with the GSD-rich region. This correspondence supports the miscut-dependent growth-mode transition proposed by Lin *et al*. [34].

In near-on-axis regions with an insufficient or nonuniform supply of surface steps, the nominal (001) growth front may locally transition from step-flow to three-dimensional faceted growth, producing a $(\bar{1}02)$ basal facet bounded by (100) sidewalls. Once formed, this faceted sector can persist and extend along [010]. Its impingement on the surrounding (001) growth region can produce the irregular terminal boundary observed by SEM. The confinement of the planar defects to the GSD termination boundaries indicates that they formed through imperfect accommodation during the impingement of the faceted and (001) growth regions. Their localized nature and weak diffraction contrast explain why they were difficult to detect by transmission XRT.

This mechanism differs from that proposed by Sdoeung *et al*. [33], who associated GSD formation with substrate dislocation clusters. In the present study, however, no one-to-one spatial correspondence was found between GSDs and substrate defects. Thus, local variations in the effective miscut angle are considered to initiate GSD formation, whereas the planar defects at the terminal boundaries are consequences of growth-front coalescence rather than nucleation origins.

## IV. CONCLUSIONS

In summary, the morphology and subsurface structure of GSDs in an HVPE-grown (001)-oriented β-$Ga_2O_3$ homoepitaxial layer were investigated using laser microscopy, SEM, transmission XRT, and cross-sectional TEM. The GSDs extended several millimeters along [010] and consisted predominantly of relatively flat $(\bar{1}02)$ basal facets and steep

(100) sidewalls. At their terminal ends, these faceted regions intersected the surrounding (001) growth surface, forming irregular, zigzag boundaries. The wafer-scale distribution of GSDs showed a spatial correspondence with variations in wafer curvature, suggesting that local surface orientation and the associated step supply influence their formation.

Careful spatial alignment of the surface morphology with transmission XRT images revealed no one-to-one correspondence between GSDs and lattice defects in the substrate or epilayer. Within the spatial resolution and contrast sensitivity of XRT, no evidence was found that substrate dislocations directly nucleate GSDs. Cross-sectional TEM observations instead revealed planar defects localized beneath the terminal boundaries of three independently examined GSDs, whereas no such defects were observed in specimens extracted from the middle of GSDs. Under two-beam diffraction conditions, these defects exhibited α-fringe contrast characteristic of inclined translational planar defects, with the dominant segments assigned to the $(1\bar{2}1)$ plane.

These results suggest that variations in local surface orientation and step supply may promote three-dimensional faceted growth, resulting in persistent $(\bar{1}02)$/(100) growth sectors. The localized planar defects are likely generated when these faceted sectors coalesce with the surrounding (001) growth region. Accordingly, the planar defects are interpreted as consequences of GSD formation and termination rather than as their nucleation origins. Control of wafer curvature and surface step supply may therefore provide a route toward suppressing GSD formation in as-grown (001)-oriented $\beta$-$Ga_2O_3$ epilayers.

## Supplementary Material

See the supplementary material for an unannotated whole-wafer laser microscopy image (Figure S1), SEM images of the GSD boundary at P4 before final FIB thinning (Figure S2), detailed two-beam TEM images of the planar defect at P6 (Figure S3), two examples showing the absence of planar defects beneath the middle portions of GSDs (Figure S4), and an XRD-derived wafer-curvature map (Figure S5).

## Acknowledgments

This study was financially supported by (1) New Energy and Industrial Technology Development Organization (NEDO) project, No. JPNP22007; (2) JSPS KAKENHI Grant No. 20K05355, 23H01872, and 23K17356 Japan. The authors are grateful to Mr. Y. Yamashita, Dr. K. Sasaki, and colleagues at NCT for preparing the samples. The synchrotron XRT observations were performed at BL24XU of SPring-8 with approval from the Japan Radiation Research Institute (Proposal Nos. 2024A3055, 2024B3055) and with the approval of RIKEN (Proposal Nos. 2025R3304, 2026R3303). The authors gratefully acknowledge Prof. Y. Tsusaka for his assistance with the XRT experiments and Ms. M. Shimizu for her assistance with the SEM and TEM observations. YY thanks Dr. T. Oshima of NIMS for many helpful discussions.

## AUTHOR DECLARATIONS

### Declaration of competing interests

The authors declare that they have no known competing financial interests or personal relationships that could have appeared to influence the work reported in this paper.

### Generative AI

Not used in the manuscript preparation process.

### Author contributions

Yongzhao Yao: conceptualization (lead); writing original draft (lead); data curation (lead); investigation (lead); writing review and editing (lead).

Daiki Katsube: investigation (equal); writing review and editing (equal).

Hirotaka Yamaguchi: investigation (equal); writing review and editing (equal).

Yukari Ishikawa: writing review and editing (equal); funding acquisition (lead).

All authors have approved the manuscript.

### Data Availability

Raw data were generated at the synchrotron facility SPring-8 and Japan Fine Ceramics Center. The data that support the findings of this study are available within the article and its supplementary material.

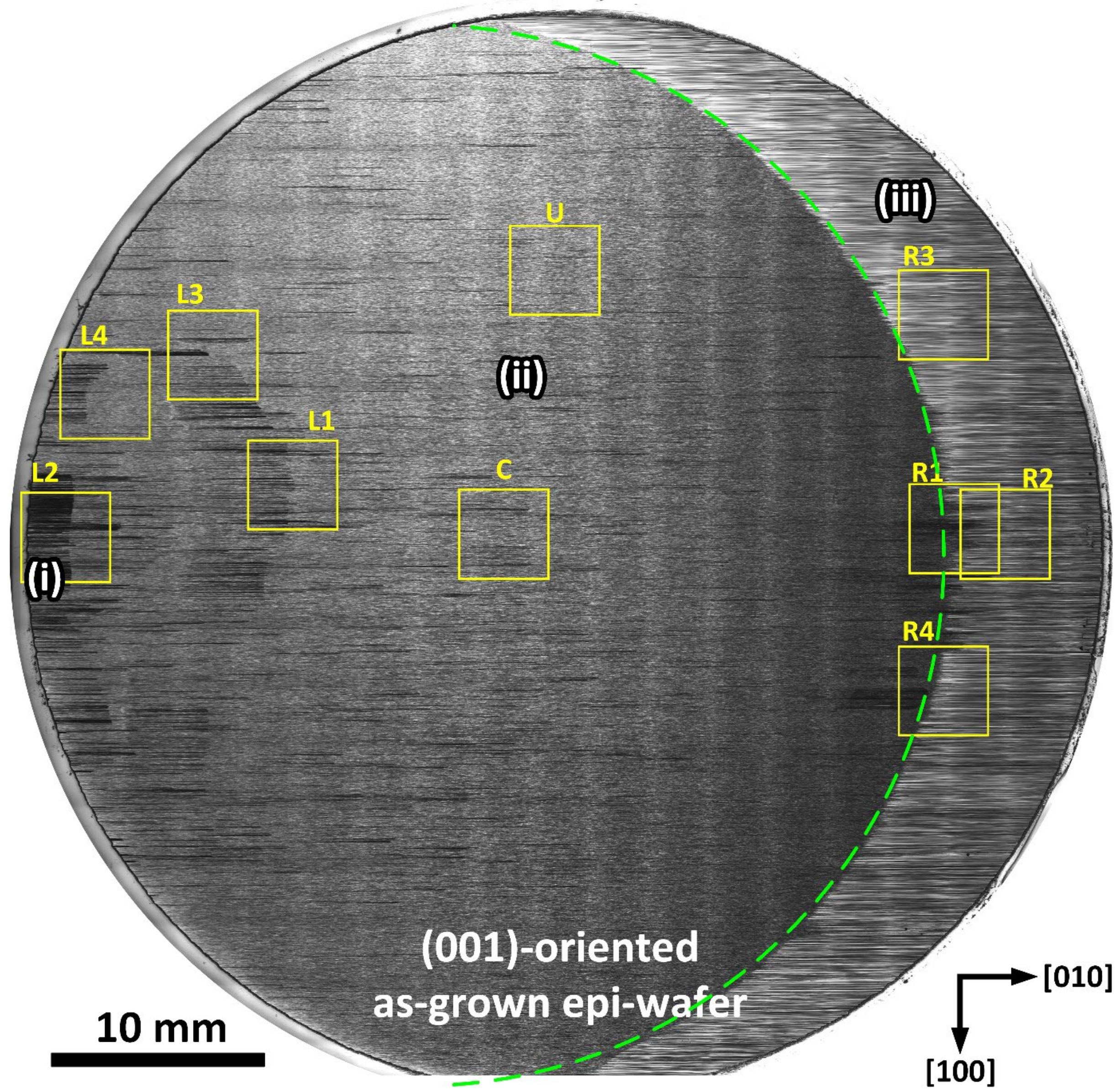

**Figure 1.** Laser microscopy image showing the surface morphology of the entire as-grown (001)-oriented β-Ga2O3 epi-wafer. The surface is classified into three regions according to image contrast: (i) a dark, GSD-rich region; (ii) an intermediate-contrast region predominantly covered with small-angle facets; and (iii) a bright region containing fewer GSDs and small-angle facets. The yellow rectangles indicate the representative areas examined in Figure 2. The green dashed line marks the boundary between regions (ii) and (iii).

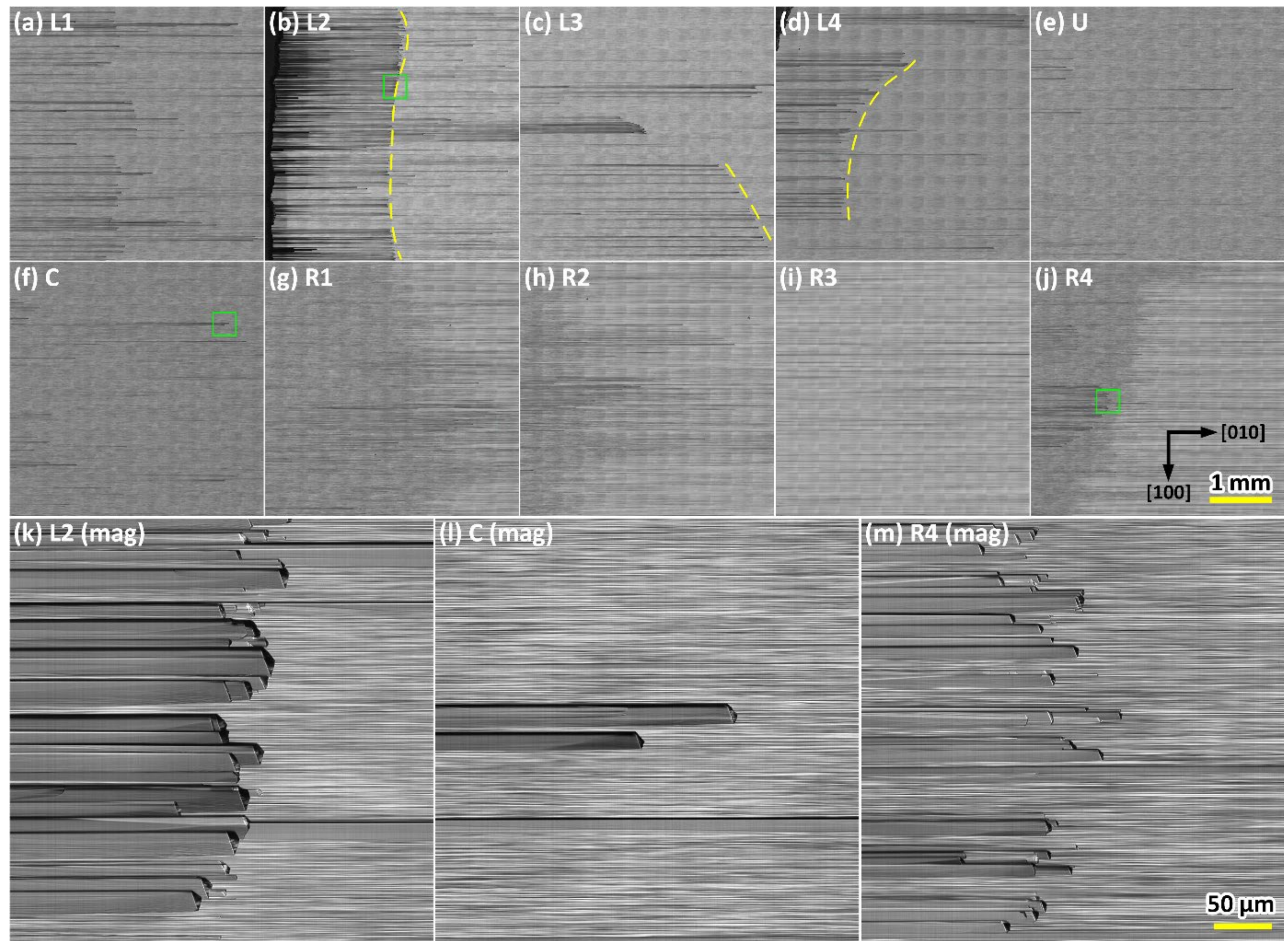

**Figure 2.** Magnified laser microscopy images of the representative areas indicated in Figure 1: (a) L1, (b) L2, (c) L3, (d) L4, (e) U, (f) C, (g) R1, (h) R2, (i) R3, and (j) R4. The yellow dashed lines in (b)–(d) indicate the termination fronts formed by groups of GSDs. (k)–(m) Higher-magnification images of the regions enclosed by the green rectangles in (b), (f), and (j), respectively. Panels (a)–(j) share the 1 mm scale bar shown in (j), whereas panels (k)–(m) share the 50 µm scale bar shown in (m).

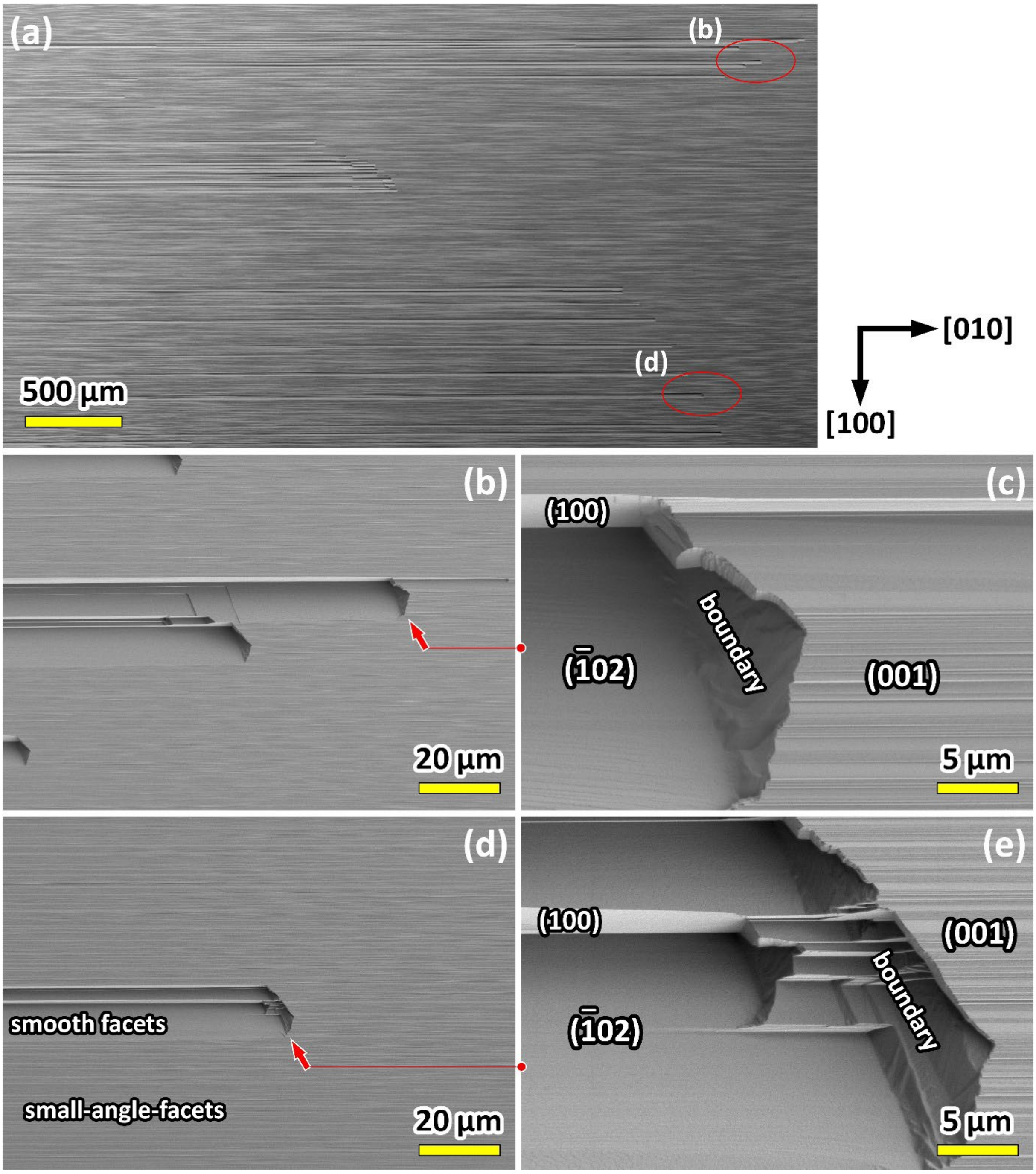

**Figure 3.** Plan-view SEM images showing the surface morphology of GSDs in area L3. (a) Overview image showing GSDs elongated along [010]. The red circles indicate the regions enlarged in (b) and (d). (b) and (d) Magnified images of representative GSD terminal regions. (c) and (e) Further-magnified images of the terminal boundaries indicated by the red arrows in (b) and (d), respectively. The GSD interiors consist predominantly of smooth ($\bar{1}$02) basal facets and (100) sidewalls, whereas the surrounding nominal (001) surface is covered with small-angle facets.

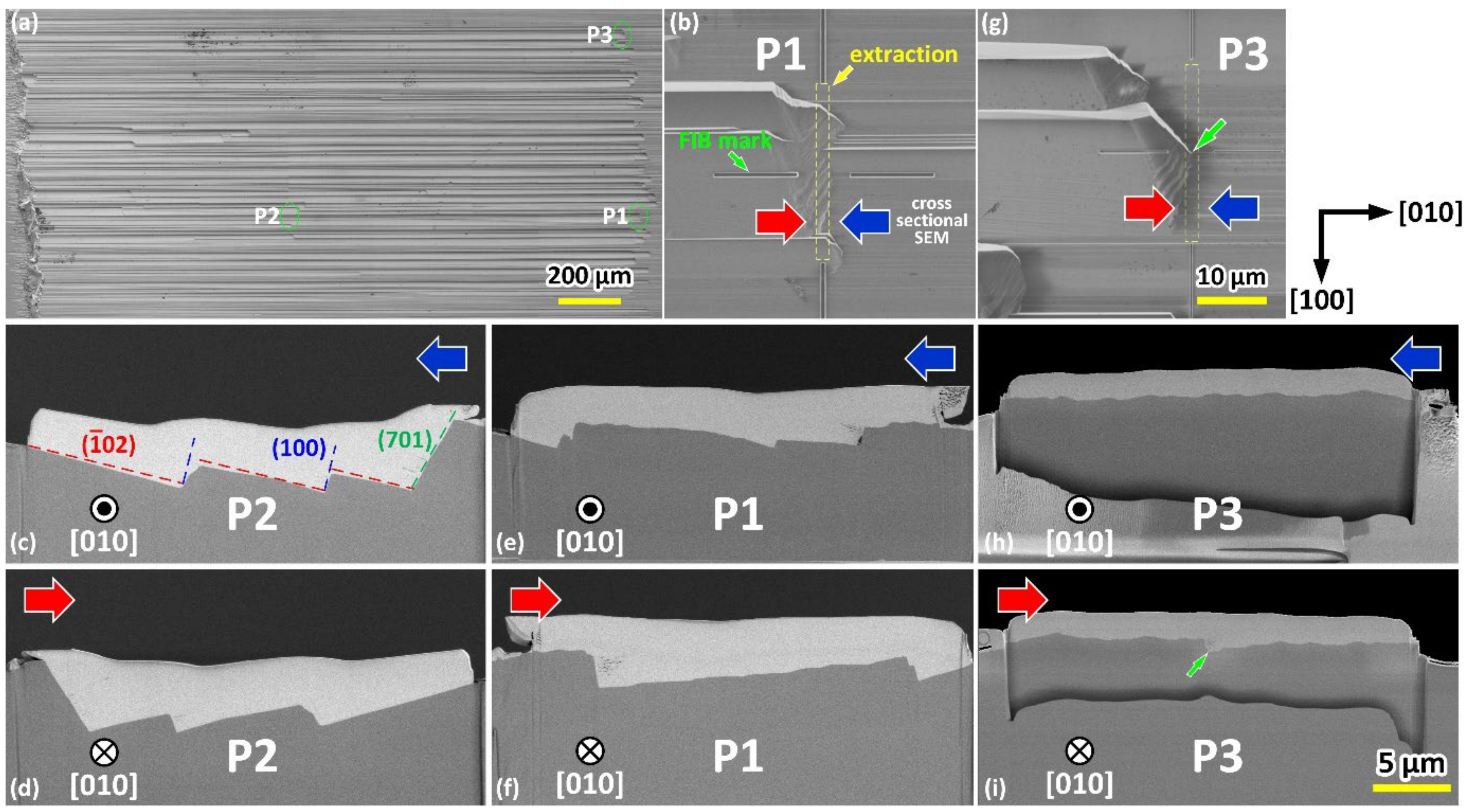

**Figure 4.** Cross-sectional SEM analysis of GSD morphology. (a) Plan-view SEM image of densely distributed GSDs in area L2. P1 and P3 are located at the terminal boundaries of two GSDs, whereas P2 is located in the middle of the same GSD as P1. (b) FIB extraction position at P1. The yellow dashed lines indicate the extraction position, and the blue and red arrows indicate viewing directions toward [010] and along [010], respectively. (c) and (d) Cross-sectional SEM images of P2 viewed from opposite directions. (e) and (f) Corresponding opposite views of the terminal specimen P1. (g) FIB extraction position at P3. (h) and (i) Opposite views of P3. Panels (b) and (g) share the 10 µm scale bar shown in (g), whereas panels (c)–(f), (h), and (i) share the 5 µm scale bar shown in (i).

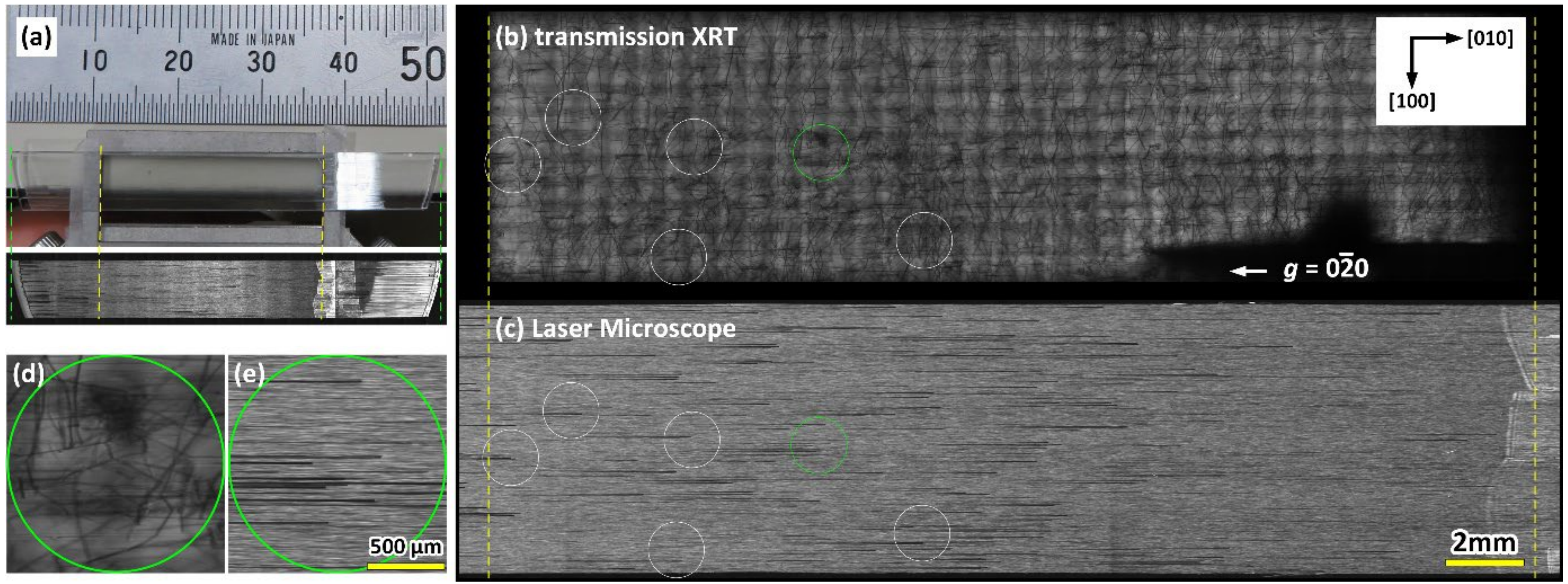

**Figure 5.** Spatial comparison between GSDs and lattice defects observed by transmission XRT. (a) Photograph of the strip-shaped specimen (top) and a laser microscopy overview of its surface (bottom). The region between the yellow dashed lines was analyzed in (b) and (c). (b) Transmission XRT image acquired with $\boldsymbol{g} = 0\bar{2}0$. (c) Laser microscopy image of the same area. The white and green circles indicate corresponding positions in the two images. (d) and (e) Magnified XRT and laser microscopy images, respectively, of the region marked by the green circles in (b) and (c).

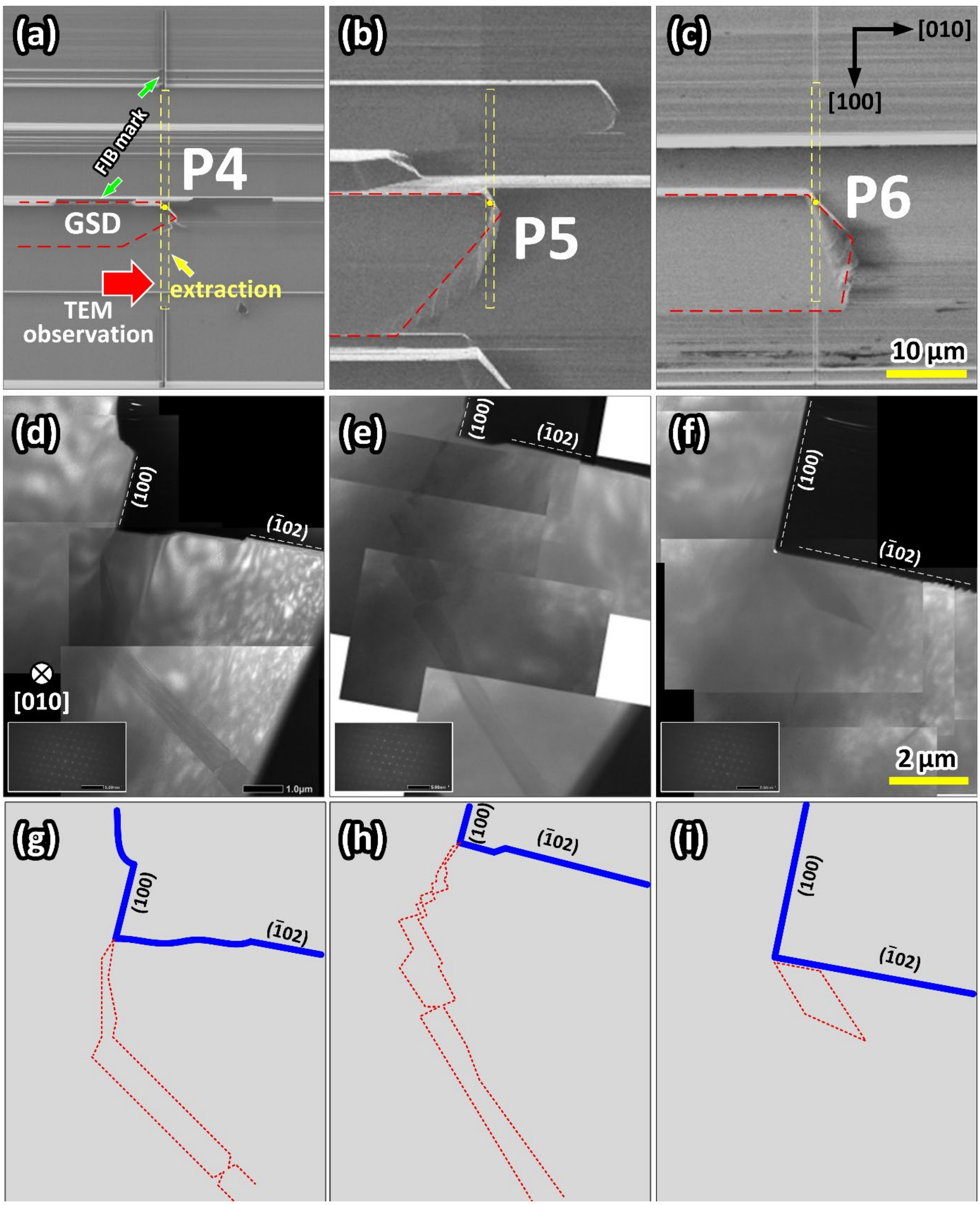

**Figure 6.** Cross-sectional TEM observations of defects beneath the terminal boundaries of three GSDs. (a)–(c) Plan-view SEM images acquired during FIB preparation of specimens P4, P5, and P6, respectively. The yellow dashed rectangles indicate the extraction positions. (d)–(f) Stitched low-magnification cross-sectional TEM images showing the entire defects beneath P4, P5, and P6, respectively. The insets show the corresponding selected-area electron diffraction patterns. (g)–(i) Schematic representations of the observations in (d)–(f), respectively.

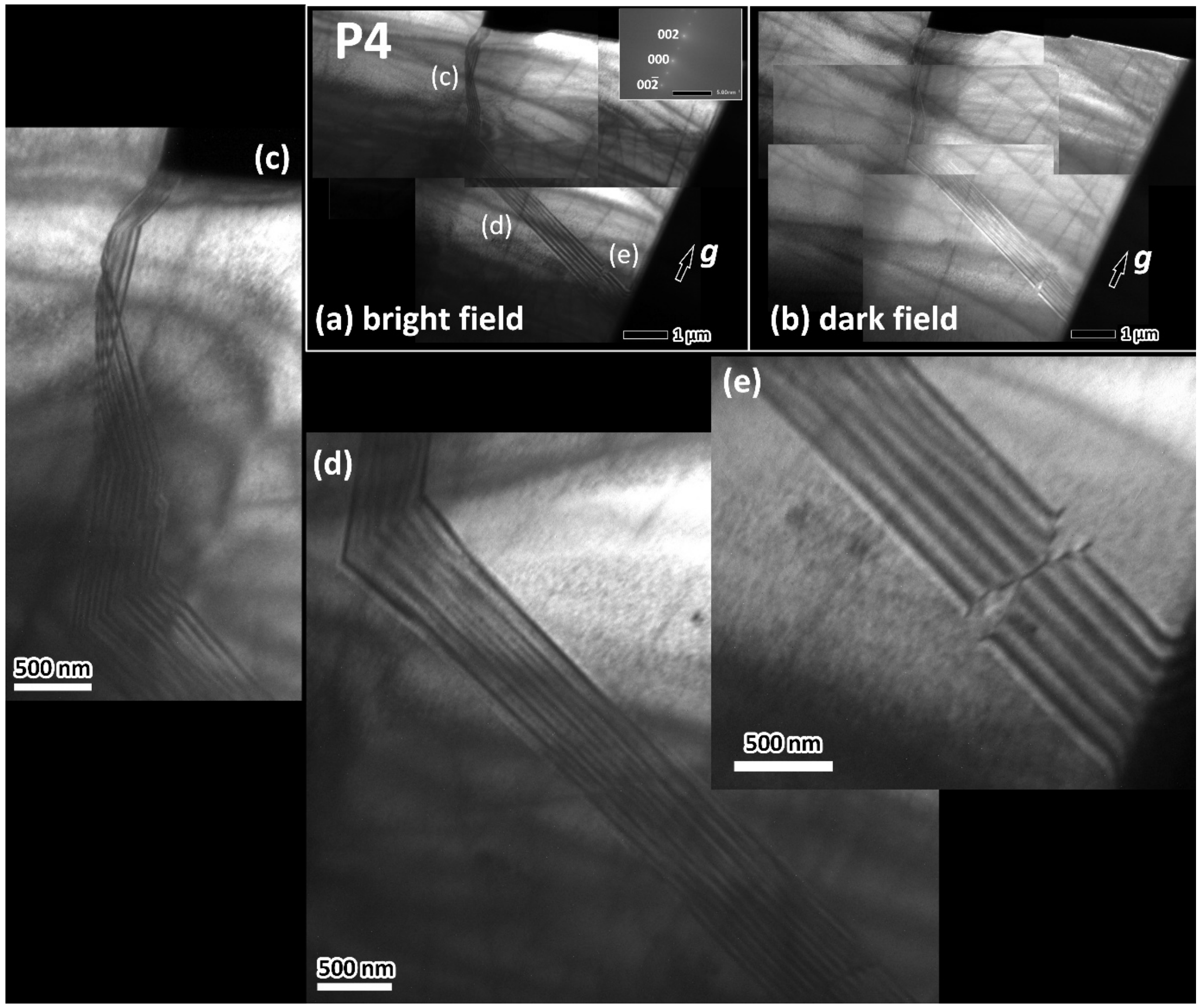

**Figure 7.** Diffraction-contrast TEM observations of the planar defect at P4. (a) Stitched bright-field TEM image acquired under a two-beam condition with $\boldsymbol{g}$ = 002. The inset shows the corresponding diffraction condition. The positions labeled (c)–(e) indicate the regions enlarged below. (b) Corresponding dark-field TEM image acquired using the same reflection. (c)–(e) Magnified bright-field images showing characteristic regions of the defect.

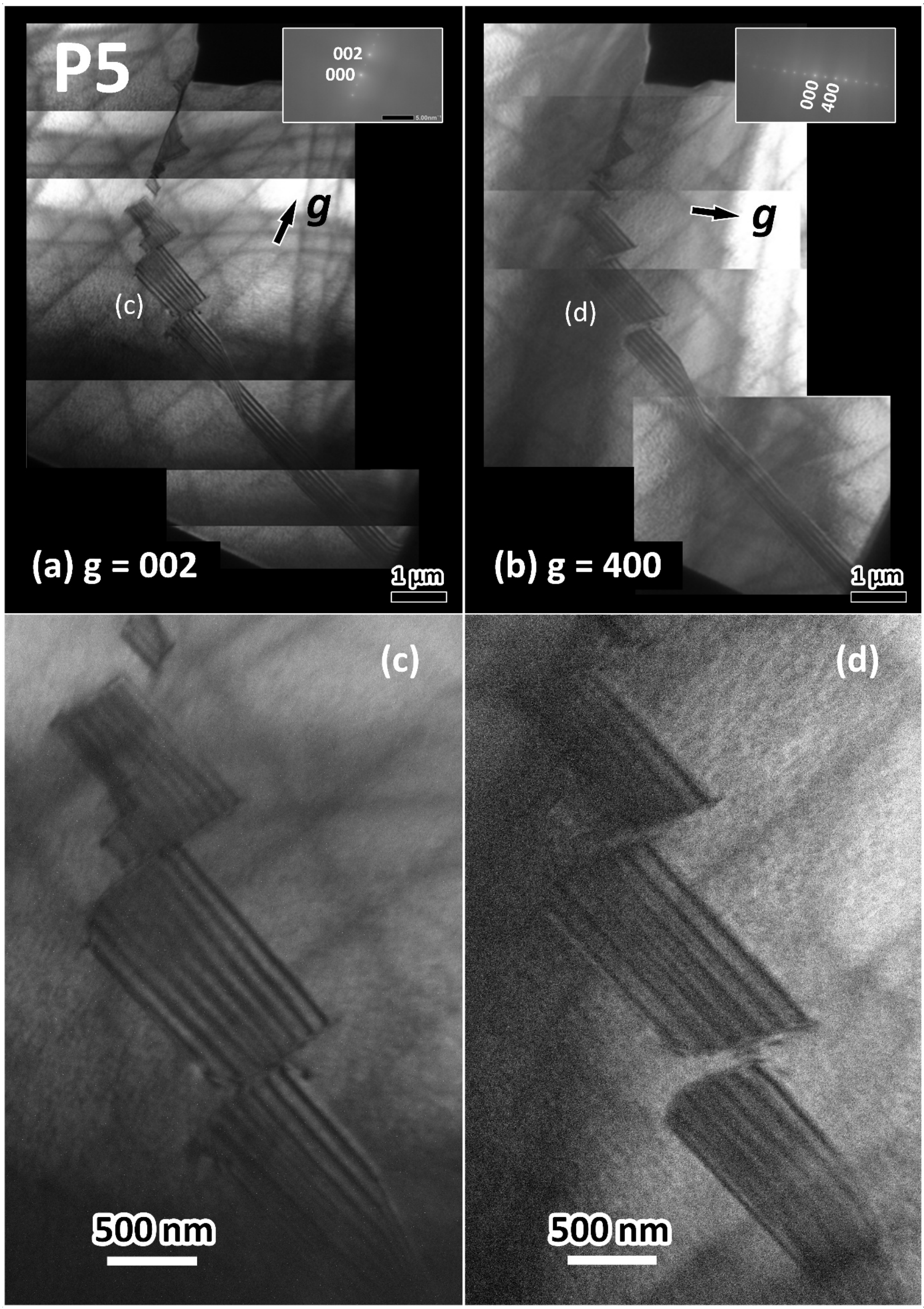

**Figure 8.** Diffraction-contrast TEM observations of the planar defect at P5. (a) and (b) Stitched bright-field TEM images acquired under two-beam conditions with ***g*** = 002 and ***g*** = 400, respectively. The insets show the corresponding diffraction conditions. (c) and (d) Magnified images of the central part of the defect obtained with ***g*** = 002 and ***g*** = 400, respectively.

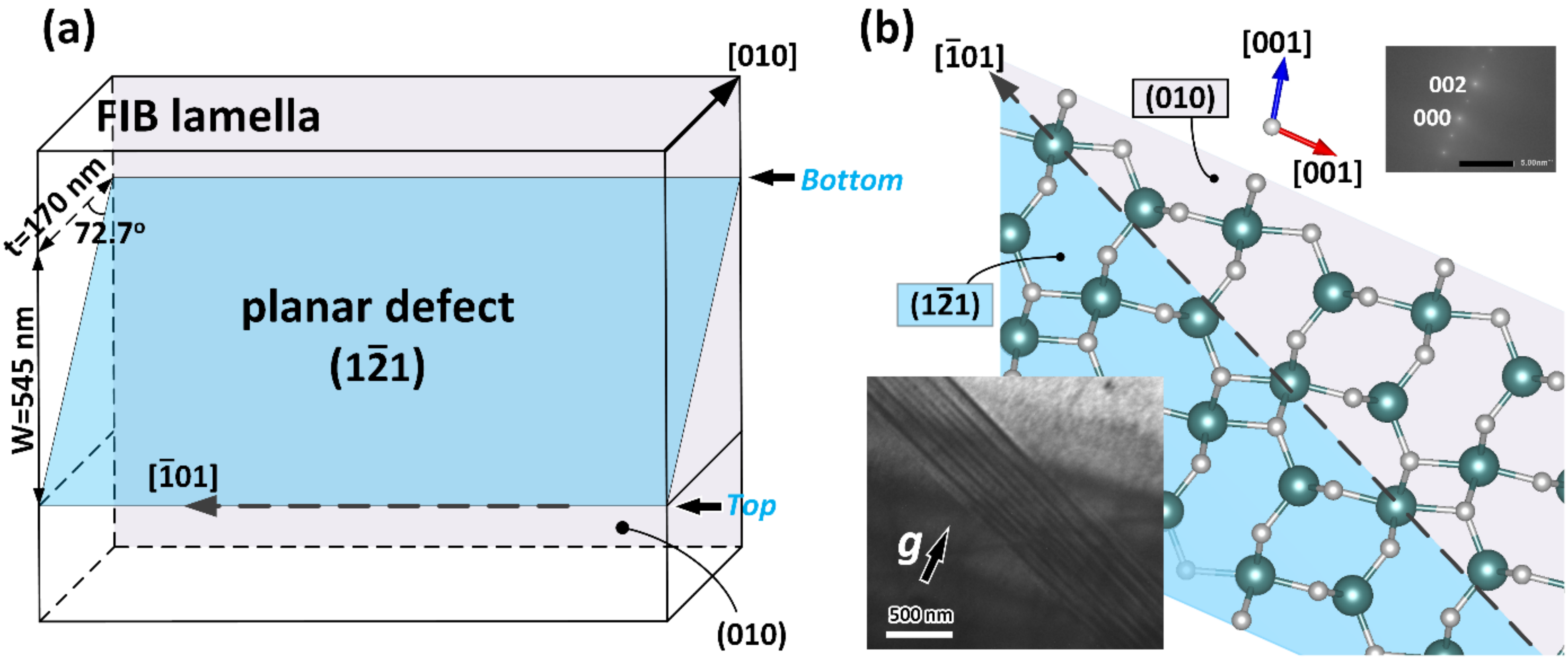

**Figure 9.** Geometrical analysis of the dominant planar-defect segments observed at P4 and P5. (a) Schematic illustration of a planar defect intersecting the top and bottom surfaces of the FIB lamella. A lamella thickness of approximately 170 nm and a projected defect width of approximately 545 nm yield an inclination angle of 72.7°, consistent with a (1$\bar{2}$1) habit plane. The dashed line indicates the [$\bar{1}$01] trace of the planar defect on the (010) lamella surface. (b) Projection of the β-$Ga_2O_3$ crystal structure showing the (1$\bar{2}$1) defect plane, the (010) lamella surface, and their intersection along [$\bar{1}$01], viewed along [010]. The TEM image shows the experimentally observed fringe direction at P4, and the diffraction pattern indicates the corresponding observation condition.

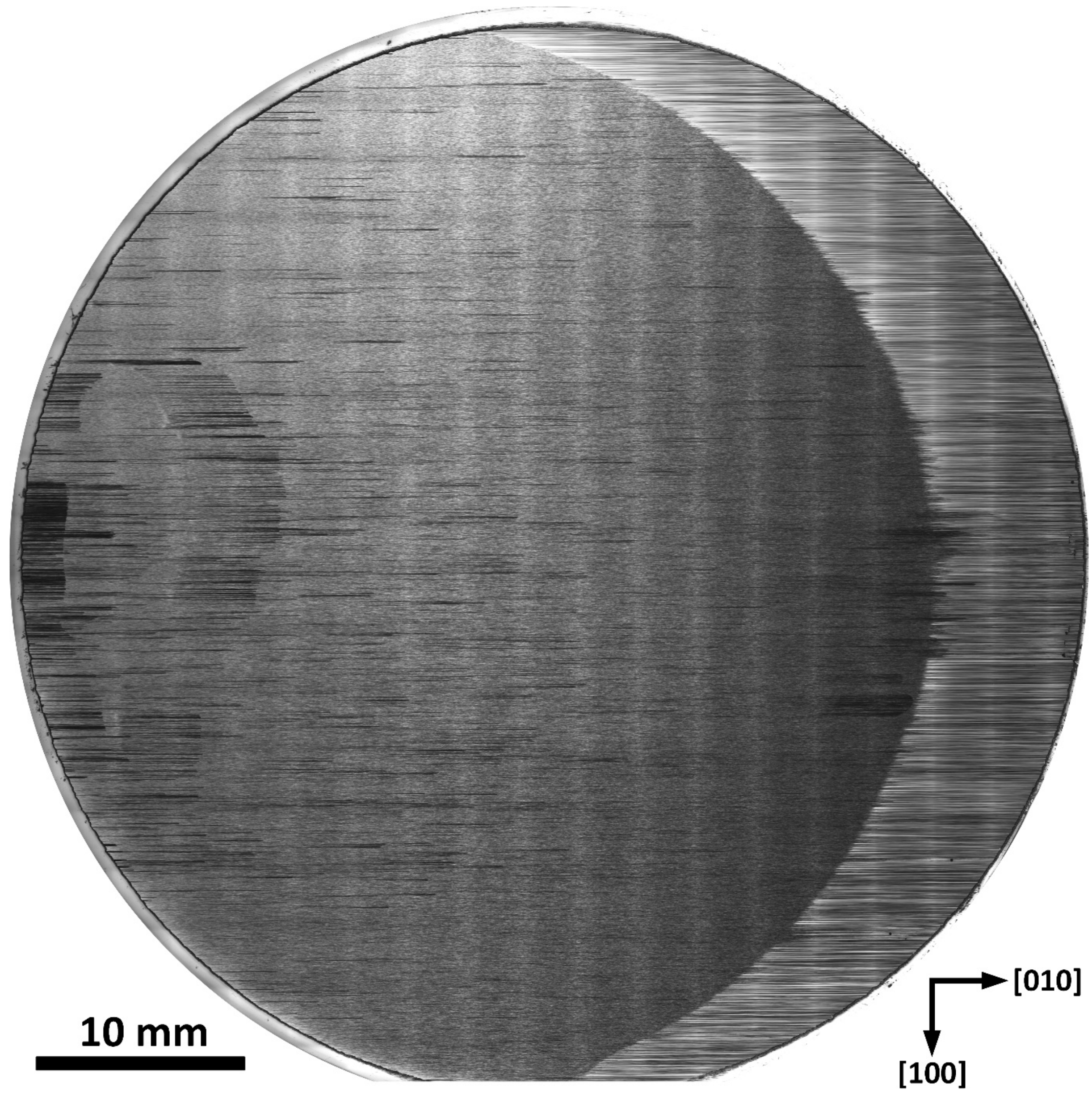

**Figure S1.** Unannotated laser microscopy image showing the surface morphology of the entire as-grown 2-inch epi-wafer.

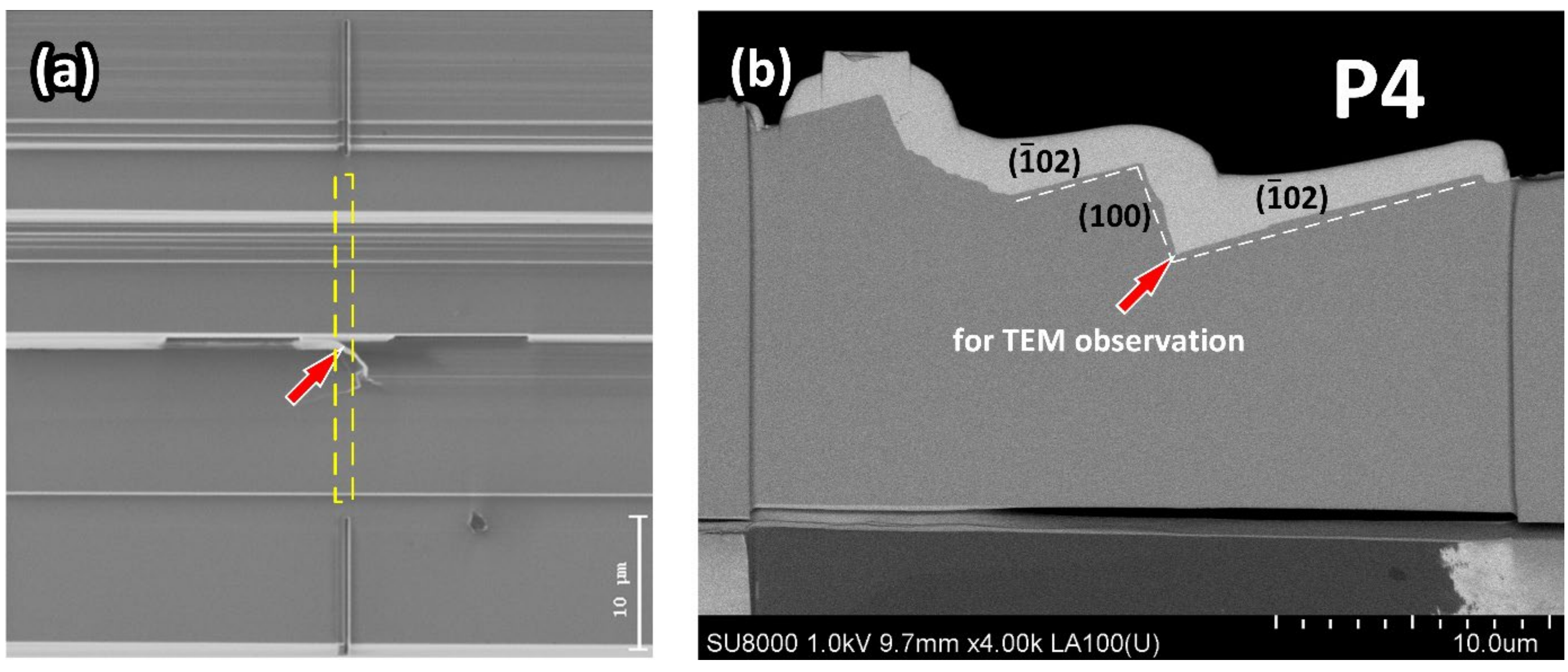

**Figure S2.** SEM observations of the GSD boundary at P4 before final FIB thinning. (a) Plan-view SEM image acquired during FIB preparation. The yellow dashed lines indicate the position of the cross-sectional specimen. (b) Cross-sectional SEM image of P4 before final thinning. The white dashed lines trace the successive ($\bar{1}02$), (100), and ($\bar{1}02$) facets, and the red arrow indicates the corner selected for TEM observation.

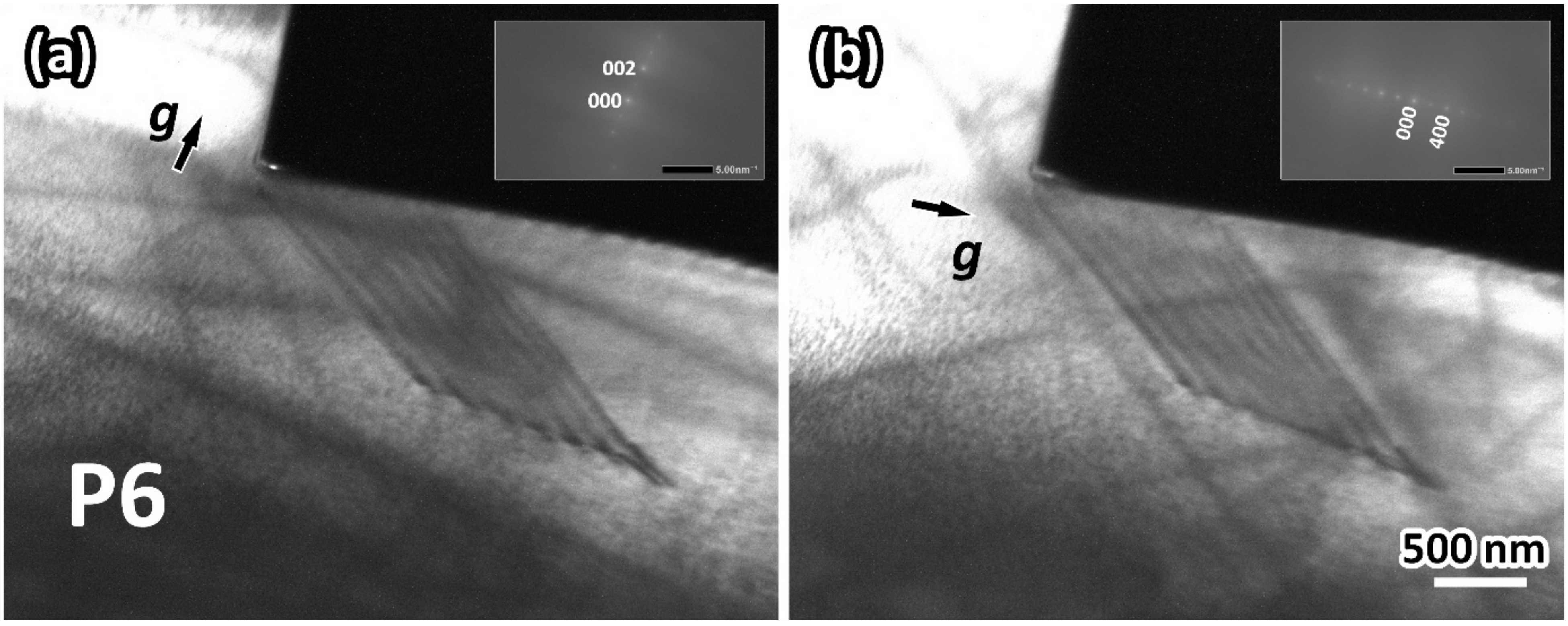

**Figure S3.** Bright-field TEM images of the planar defect at P6 acquired under two-beam conditions using (a) ***g*** = 002 and (b) ***g*** = 400. The corresponding diffraction conditions are shown in the insets.

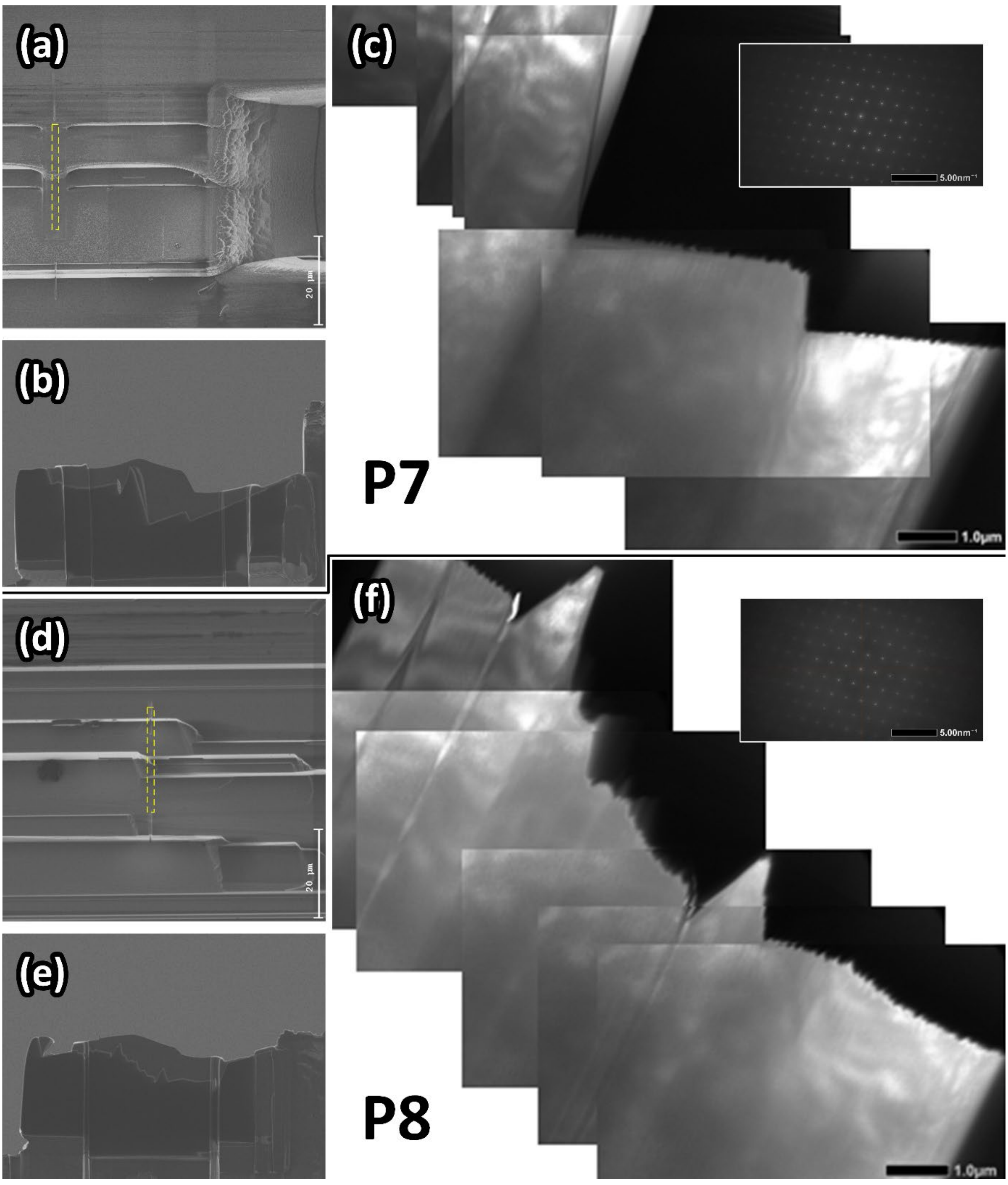

**Figure S4.** Cross-sectional observations of specimens extracted from the middle portions of two GSDs, designated P7 and P8. (a) and (d) Plan-view SEM images showing the extraction positions at P7 and P8, respectively; the yellow dashed rectangles indicate the sampled regions. (b) and (e) SEM images of the corresponding specimens after FIB preparation. (c) and (f) Stitched low-magnification TEM images of P7 and P8, respectively, with the corresponding selected-area electron diffraction patterns shown in the insets. No planar defects were detected beneath the middle portion of either GSD.

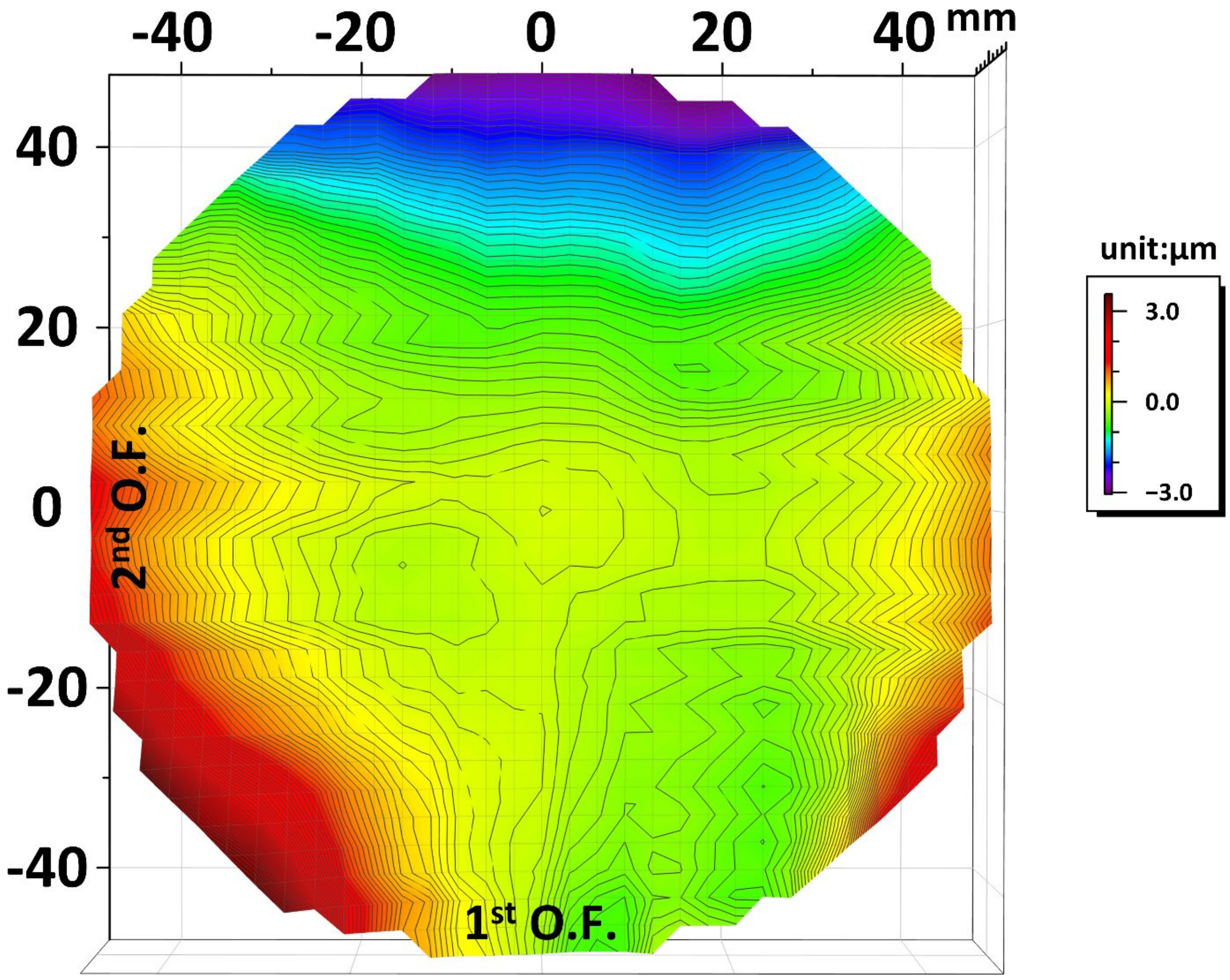

**Figure S5.** XRD-derived curvature map of the 2-inch wafer, reconstructed from spatial variations in the local normal direction of the (001) lattice plane. The positions of the primary and secondary orientation flats are indicated. The colors and contour lines represent the relative out-of-plane displacement of the wafer surface in micrometers.